\documentclass{article}

\begin{document}

\title{Is the active gravitational mass of a charged body distance-dependent?}
\author{Vesselin Petkov \\
Physics Department, Concordia University\\
1455 de Maisonneuve Boulevard West \\
Montreal, Quebec H3G 1M8\\
vpetkov@alcor.concordia.ca}
\date{12 April 2001}
\maketitle

\begin{abstract}
It appears to follow from the Reissner-Nordstr\o m solution of Einstein's
equations that the charge of a body reduces its gravitational field. In a
recent note Hushwater offered an explanation of this apparent paradox. His
explanation, however, raises more questions than solves since it implies
that the active gravitational mass of a charged body is distance-dependent
and therefore is not equal to its inertial mass.
\end{abstract}

\noindent As discussed by Hushwater \cite{hushwater} the Reissner-Nortstr\o %
m solution of Einstein's equations for a charged body of mass $M$ and charge
$Q$

\begin{equation}
ds^{2}=-\left( 1-\frac{2GM}{c^{2}r}+\frac{GQ^{2}}{c^{4}r^{2}}\right)
c^{2}dt^{2}+\left( 1-\frac{2GM}{c^{2}r}+\frac{GQ^{2}}{c^{4}r^{2}}\right)
^{-1}dr^{2}+r^{2}\left( d\theta ^{2}+\sin ^{2}\theta d\phi ^{2}\right)
\label{rn}
\end{equation}
may lead one to the conclusion that the body's gravitational filed is
reduced by its charge. This becomes obvious if the Newtonian limit of
general relativity is considered. In that limit the metric tensor of curved
spacetime $g_{\alpha \beta }$ can be represented by the metric tensor of
flat spacetime $\eta _{\alpha \beta }$ and another ''perturbation'' tensor $%
h_{\alpha \beta }$ whose components are much less than unity (since they are
proportional to $c^{-n}$ where $n\geq 2$)

\[
g_{\alpha \beta }=\eta _{\alpha \beta }+h_{\alpha \beta }.
\]
In the Newtonian limit of Reissner-Nortstr\o m metric

\[
g_{00}=1+h_{00}
\]
where

\[
h_{00}=-\frac{2GM}{c^{2}r}+\frac{GQ^{2}}{c^{4}r^{2}}.
\]
The equation of motion of a non-relativistically moving test particle in
terms of $h_{00}$ is then

\[
\frac{d^{2}\mathbf{r}}{dt^{2}}=\frac{c^{2}}{2}\nabla h_{00}
\]
or

\begin{equation}
\frac{d^{2}\mathbf{r}}{dt^{2}}=-\nabla \left( \frac{GM}{r}-\frac{GQ^{2}}{%
2c^{2}r^{2}}\right) .  \label{dr}
\end{equation}
As seen from (\ref{dr}) it appears that the charge $Q$ of the body reduces
its gravitational field since the second term in the parentheses is
subtracted from the gravitational potential $GM/r$. The problem with such a
conclusion is that it contradicts the very foundations of relativity
according to which the electric field of the body must, in fact, increase
its gravitational field since the electric field possesses energy and
therefore mass.

Hushwater claims to have resolved this apparent paradox by making use of the
concept ''total mass inside a radius $r$'' felt by a test particle at a
distance $r$ from the body's center

\begin{equation}
M\left( r\right) =M-\frac{1}{8\pi c^{2}}\int |\mathbf{E}|^{2}d^{3}x
\label{Mr}
\end{equation}
where $M$ is the total mass of the charged body consisting of its ordinary
mass and the \emph{whole} electromagnetic mass that corresponds to the
energy of the body's electric field $\mathbf{E}$. The integration in the
second term in (\ref{Mr}) is taken over the space outside a sphere $S(r)$ of
radius $r$ and therefore that term is the \emph{part} of the electromagnetic
mass that is stored in the body's electric field occupying the space outside
$S(r)$. In such a way the mass $M\left( r\right) $ comprises the ordinary
mass of the body and only that part of its electromagnetic mass that
corresponds to the energy of the electric field inside the sphere $S(r)$. As

\[
\mathbf{E}=-\nabla \frac{Q}{r}=\frac{Q}{r^{2}}\mathbf{n}
\]
where $\mathbf{n}=\mathbf{r}/r$, the second term in (\ref{Mr}) is equal to $%
Q^{2}/2c^{2}r$. Therefore for (\ref{Mr}) one obtains

\begin{equation}
M\left( r\right) =M-\frac{Q^{2}}{2c^{2}r}.  \label{M(r)}
\end{equation}
The classical equation of motion of a particle in a gravitational potential $%
GM\left( r\right) /r$ caused by the mass $M\left( r\right) $ is

\begin{equation}
\frac{d^{2}\mathbf{r}}{dt^{2}}=-\nabla \frac{GM\left( r\right) }{r}.
\label{dr2}
\end{equation}
The substitution of (\ref{M(r)}) in (\ref{dr2}) gives (\ref{dr}). This
result is regarded by Hushwater as a proof that the apparent paradox
disappears if both the ordinary and electromagnetic mass of a charged body
are taken into account and the mass $M\left( r\right) $ inside a sphere $S(r)
$ is used.

This resolution of the paradox, however, comes at too high a price since it
is based on two implicit assumptions none of which seems to be correct.

1. The mass $M\left( r\right) $ is implicitly regarded as the \emph{active
gravitational mass} of the charged body, which is felt by a test particle
placed at a distance $r$ from the body's center. And indeed, as seen from (%
\ref{dr2}) it is the mass $M\left( r\right) $ that gives rise to
the gravitational potential $GM\left( r\right) /r$. This means,
however, that the active gravitational mass of a charged body is
\emph{distance-dependent}. Leaving aside the question of what
that may mean, it is obvious that a distance-dependent active
gravitational mass is not equal to the body's inertial mass,
which consists of its ordinary mass and its \emph{entire}
electromagnetic mass; $M\left(r\right)$ coincides with the
inertial mass of the charged body only when $r\rightarrow \infty
$. An assumption that the active gravitational mass of a charged
particle is not equal to its inertial mass is not justified since
there exists no evidence that the equivalence principle is
violated in the case of charged particles \cite{equiv}.

2. Regarding the mass $M\left( r\right) $ as the source of the gravitational
potential $GM\left( r\right) /r$ in (\ref{dr2}) implies that $M\left(
r\right) $ defines some metric. To find that metric one can substitute the
expression for $M\left( r\right) $

\[
M-\frac{Q^{2}}{2c^{2}r}=M\left( r\right)
\]
from (\ref{M(r)}) in (\ref{rn}). The result is

\begin{equation}
ds^{2}=-\left( 1-\frac{2GM\left( r\right) }{rc^{2}}\right)
c^{2}dt^{2}+\left( 1-\frac{2GM\left( r\right) }{rc^{2}}\right)
^{-1}dr^{2}+r^{2}\left( d\theta ^{2}+\sin ^{2}\theta d\phi ^{2}\right)
\label{ds}
\end{equation}
which is the Schwarzschild metric in the case of a body of active
gravitational mass $M\left( r\right) $. In other words, the
second implicit assumption is that the Reissner-Nortstr\o m
metric (\ref{rn}) can be directly obtained from the Schwarzschild
metric (\ref{ds}) if the expression for the mass $M\left(
r\right) $ is substituted in (\ref{ds}). However, it is obviously
incorrect to use $M\left( r\right) $ in the Schwarzschild
solution since it is a vacuum solution $\left( T_{\alpha \beta
}=0\right) $ whereas in the case of a charged body $T_{\alpha
\beta }\neq 0$.

If the equivalence principle strictly holds for charged
particles, then the paradox that the charge of a body reduces its
gravitational field according to the Reissner-Nortstr\o m
solution remains. And if no other explanation of that paradox is
found the only way out of this situation seems to be to assume
that the active gravitational mass of a charged body is indeed
distance-dependent and therefore is not equal to its inertial
mass \cite{test}. This would mean that the Reissner-Nortstr\o m
solution of Einstein's equations does follow from the
Schwarzschild solution if the expression for the mass $M(r)$ is
substituted in (\ref{ds}). Then a justification for such an
illegal at least at first glance operation might be the
following. A test particle at a distance $r$ from the charged
body's center does not feel the effect of the electromagnetic
mass corresponding to the body's electric field outside the
sphere $S(r)$ since it cancels out exactly. In this sense the
space outside the sphere $S(r)$ will appear "empty" to the test
particle.


\begin{thebibliography}{9}
\bibitem{equiv} Strictly speaking, the equivalence principle
states that there is an equivalence between inertial and passive
gravitational mass. However, the active and passive gravitational
masses are presisely equal (at least in the Newtonian
gravitational theory) as seen from the fact that Newton's
gravitational law $F=GMm/r^{2}$ (where $M$ and $m$ are the active
gravitational masses of two bodies $1$ and $2$, respectively) can
be written as $F=mg$, where $g=GM/r^{2}$ is the gravitational
acceleration and $m$ is the passive gravitational mass of body
$2$.
\bibitem{test} With appropriate values of the charge $Q$ and the
distance $r$ as seen from (\ref{M(r)}) a violation of the
equivalence principle is in principle detectable - see
\cite{braginsky}. However, testing the equivalence of the active
gravitational mass of a charged body and its inertial mass may be
a real challenge due to the presense of the body's electric field.
\end{thebibliography}
\end{document}